# Structure, dynamics and phase behavior of short rod inclusions dissolved in a colloidal membrane

Mahsa Siavashpouri[1], Prerna Sharma[1,2], Jerome Fung[1,3], Michael. F. Hagan[1], Zvonimir Dogic[1,4]
[1]Department of Physics, Brandeis University, Waltham, MA 02454, USA
[2]Department of Physics, Indian Institute of Science, Bangalore 560012, India
[3]Department of Physics, Ithaca College, Ithaca, NY 14850, USA
[4]Department of Physics, University of California, Santa Barbara, CA 93106, USA

**Abstract:** Inclusions dissolved in an anisotropic quasi-2D membrane acquire new types of interactions that can drive assembly of complex structures and patterns. We study colloidal membranes composed of a binary mixture of long and short rods, such that the length ratio of the long to short rods is approximately two. At very low volume fractions, short rods dissolve in the membrane of long rods by strongly anchoring to the membrane polymer interface. At higher fractions, the dissolved short rods phase separate from the background membrane, creating a composite structure comprised of bilayer droplets enriched in short rods that coexist with the background monolayer membrane. These results demonstrate that colloidal membranes serve as a versatile platform for assembly of soft materials, while simultaneously providing new insight into universal membrane-mediated interactions.

**Introduction:** Colloids, proteins and nanoparticles dissolved in bulk isotropic fluids interact by well-studied intermolecular forces that include steric exclusions, electrostatic repulsions, the hydrophobic effect, and van der Waals interactions[1]. In comparison, particles dissolved in anisotropic environments or confined on surfaces or interfaces can acquire more complex interactions and thus exhibit very different behaviors. For example, experiments have revealed exceedingly complex interactions and assembly pathways of colloids or nano-particles dissolved in anisotropic liquid crystals[2-4] or confined on oil-water interfaces[5-9]. Lipid bilayers provide an even more complex environment for self-assembly. Particles dissolved in a lipid bilayer simultaneously experience a liquid crystalline environment due to ordering of the hydrophobic lipid chains[10], and are confined to a deformable quasi-2D plane, similar to particle-laden interfaces. Consequently, membrane-mediated interactions can drive assembly of exceedingly complex structures[11-15]. However, the nanometer length scale of conventional lipid bilayers makes studies of lipid bilayers challenging. Consequently, our knowledge of membrane-mediated interactions and assembly processes remains underdeveloped, especially when compared to 3D colloidal-liquid crystal mixtures or particles confined on 2D interfaces.

Recent experiments demonstrated a distinct pathway for assembly of 2D membrane-like structures that relies on the geometry of the constituent particles rather than their chemical heterogeneity. In the presence of non-adsorbing depleting polymers, monodisperse colloidal rods assemble into liquid-like one-rod length thick colloidal membranes[16-19]. Although colloidal membranes are more than two orders of magnitude thicker than lipid bilayers, the deformations of both systems are described by the same elastic energy. The intrinsic length scale of colloidal membranes allows for visualization of how inclusions distort the membrane structure, and for measurement of membrane mediated-interactions. For example, experiments demonstrated that chiral objects dissolved in a 2D membrane acquire long-range repulsive interactions, leading to formation of thermodynamically stable finite-sized colloidal rafts, which are micron sized liquid droplets enriched in shorter rods[20-23]. Here, we study 880 nm thick colloidal membranes, in which we dissolve rods that are approximately half the membrane thickness. We map the phase diagram of this two-component mixture uncovering rich phase behaviors. At low densities, a dislocation defect created by a rod end anchors short rods to the membrane-polymer interface. Anchored rods occasionally hop across the membrane midplane to the opposite interface. With increasing concentration, short rods phase separate from a background monolayer membrane, forming 2D liquid bilayer droplets that coexist with the background monolayer membrane. The rod asymmetry of the binary mixture we study is significantly larger than those studied previously[20, 23], demonstrating that changing rod length leads to different behaviors.

**Materials and methods:**

**Bacteriophage growth and purification:** Bacteriophages *fd-wt*, *fd*-Y21M and *litmus* were grown and purified using standard biological procedures[24]. Plasmid DNA sequence of *litmus* 38i was used to form *litmus* phagemid which was grown to large scale using M13K07 as the helper phage. Robust formation of colloidal membranes requires samples that have minimum contamination of longer rods that can sometimes be present in the sample preparation. To eliminate these samples we have used previously developed protocol[16]. Briefly, purified virus suspensions were phase-separated through the isotropic–nematic phase transition as described previously. Only the isotropic fraction, enriched in *litmus* monomers, was use for further experimentation. All three purified viruses were suspended in 135 mM NaCl and 20 mM Tris-HCl buffer (pH 8.0) to screen rod-rod electrostatic interactions. Sample polydispersity was checked using gel electrophoresis on the intact virus and on the viral DNA.

**Bacteriophage labeling:** For fluorescence microscopy, primary amines of the major coat protein of *litmus* were labeled with amine reactive fluorophore (DyLight-NHS ester 550; Thermo Fisher)[25]. There are about 1,200 labeling sites available on the virus surface. However, each virus was labeled at a low volume fraction (25 dye molecules per virus). The system phase behavior depends on the degree of labeling of the labeled rods. Labeling at lower densities (5 dye molecules per virus) slows the lateral phase separation. To visualize dynamics of single *litmus* rods, all *fd-wt* filaments were labeled with Alexa 488 (25 dye molecule per virus) and were mixed with *litmus* virus where 1 out of 10,000 *litmus* rods were labeled by DyLight550 fluorophore. To reduce photobleaching effects, we added a standard oxygen scavenging solution consisting of glucose oxidase, catalase and glucose[26].

**Sample preparation**: Bacteriophages *fd-wt* and *litmus,* as well as *fd*-Y21M and *Litmus*, were mixed at known stoichiometric ratios. The non-adsorbing polymer dextran (molecular weight 670,000 Da; Sigma-Aldrich) was added to this suspension. The final concentration of bidisperse virus mixture was 2 mg/mL. The optical density of *litmus*, *fd-wt* and *fd*-Y21M for 1 mg/mL solution is 3.84 at $\lambda$=269 nm. Final concentration of polymer varied from 20 mg/mL to 80 mg/mL, while the concentration of NaCl was adjusted to 135 mM. The samples were injected into an optical microscopy chamber that consisted of one glass slide and one coverslip attached together via a layer of unstretched parafilm. To prevent nonspecific binding of virus to the glass slide and coverslip surfaces, glass surfaces were coated with polyacrylamide brush[27]. Self-assembled structures formed in a few hours and slowly sedimented to the coverslip due to their higher density.

**Optical Microscopy:** Samples were visualized by an inverted microscope (Nikon TE-2000) equipped with a differential interference contrast (DIC) module, a fluorescence imaging module and a 2D-LC-Polscope module[28]. A high numerical aperture oil objective (100x PlanFluor NA 1.3) and a mercury halide lamp (Excite-120) were used. Images were collected with a cooled CCD cameras (Andor-Clara for DIC and LC-Polscope imaging, Andor iXon for fluorescence imaging). Fluorescently labeled *litmus* viruses were imaged using a rhodamine filter cube (excitation wavelength 532–554 nm, emission wavelength 570–613 nm). Fluorescently labeled *fd-wt* viruses were imaged using a FITC filter cube (excitation wavelength maximum at 490nm and emission wavelength maximum at 525 nm). For quantitative analysis of the fluctuating interface, the exposure time was kept at minimum (less than 20 ms) to reduce blurring effects.

**Transmission Electron Microscopy (TEM):** To visualize individual bacteriophage filaments, negative stain TEM were employed. After glow-discharging the carbon-coated TEM grids, we stained them with 2% Uranyl-Formate stain solution with 25 mM NaOH. 4 μL of sample solution at 3 nM concentration was applied on the carbon-coated side of the TEM grid. Imaging was performed using a CM12 electron microscope operated at 100 kV. Images were acquired using an AMT (Advanced Microscopy Techiques Corp., Danvers, MA) CCD system.

**Analyzing fluid-fluid interface fluctuations with fluorescence microscopy**: **Analyzing fluid-fluid interface fluctuations with fluorescence microscopy**: We describe the algorithms used to analyze the fluctuations of the fluid-fluid interfaces – *fd wt* and *litmus*, *fd*-Y21M and *litmus*, and *litmus* outer edge – shown in Fig. 7. In all cases, the principal challenge is to locate the edge of a region of fluorescently labeled rods with subpixel precision. Our technique is schematically illustrated in Fig. S1. Our analyses use custom Python code that draws heavily on the *numpy* and *scipy* libraries. A portion of a raw fluorescence image of the membrane shown in Fig. 7b is shown in Fig. S1a; the analysis of the membrane in Fig. 7a is similar. First, we calculate the magnitude of the gradient of the raw image by applying Sobel operators (Fig. S1b). This gradient magnitude image is large at the edges of the bright fluorescent region in the original image. Second, we use Canny edge detection[29] to roughly locate the fluorescent edges in the raw image to within a pixel. We briefly summarize the Canny edge detection algorithm. The original image is smoothed with a Gaussian filter, and then the horizontal and vertical components of the gradient of the smoothed image are computed using Sobel operators. A set of candidate edge pixels is then determined by choosing pixels in the magnitude of the gradient of the smoothed image that exceed a threshold value. Since the set of candidate edge pixels is usually more than one pixel thick, a non-maximum edge suppression scheme then retains only candidate edge pixels that are local maxima along the direction of the gradient at that pixel. Occasionally, small clusters of spurious pixels remain. We remove the spurious pixels by eliminating any pixels that are not directly connected to at least one pixel whose gradient magnitude exceeds a second, higher threshold value. The result is a one-pixel-thick set of edge points (Fig. S1c).

Third, we refine the edge locations. For each of the two edges (*fd*-Y21M and *litmus*, and *litmus* outer edge) in Fig. S1c, we first fit cubic splines with a smoothing parameter to the Canny edge points (Fig. S1d). The smoothing parameter allows the spline curves to not pass exactly through the Canny edge points. Then, we proceed along each rough spline curve in arc length increments of one pixel width. At each location along a rough spline curve, we examine the intensity of the gradient magnitude image along a direction perpendicular to the rough spline curve, using linear interpolation to determine the gradient magnitude at any points that do not correspond to an exact pixel location. We fit Gaussians to the resulting cut through the gradient magnitude to determine the location of the maximum to subpixel precision (Fig. S1e). This results in a set of refined edge points (Fig. S1f).

Finally, we fit a new set of cubic splines with no smoothing to the refined edge points. This refined spline curve is constrained to pass through all of the refined edge points and facilitates the computation of the tangent angles to the edge. The averages of the squares of the Fourier components of these tangent angles lead to the fluctuation spectra shown in Fig. 7c. The fluctuation spectra are not sensitive to the exact values of the empirically-determined parameters used in this algorithm (namely, the Canny thresholds, Gaussian filter width, and cubic spline smoothing parameter).

**Experimental results:** Adding non-adsorbing polymer (Dextran, MW 670,000) to a dilute isotropic solution of rods induces lateral attractive interactions leading to assembly of colloidal membranes, which are one rod-length thick liquid-like monolayers of aligned rods. We assembled colloidal membranes using a binary mixture of 880 nm long *fd-wt* and 385 nm long *litmus* bacteriophages. Both viruses organize into a cholesteric phase with left-handed twist and have a persistence length of 2.8 μm (Fig.1a)[30-33]. Bacteriophage *litmus* has a major coat protein identical to M13 virus, which differs from *fd-wt* coat protein by a single charged amino acid. Consequently, *litmus* rods have a lower surface charge than *fd-wt* rods and pack to higher densities at the same osmotic pressure[34]. In order to distinguish the two virus types, *litmus* rods were labeled with Dylight-550 (shown as yellow channel) and *fd-wt* rods were labeled with Alexa-488 (shown as blue channel) (See methods). The membrane composition is defined as $n_{mem}=N_{litmus}/N_{fd}$, where $N$ is the areal virus number density.

We first assembled membranes at a very low volume fraction of short rods ($n_{mem}$ = 3 x$10^{-4}$) and a Dextran concentration of 40 mg/mL. For these conditions all short rods were fluorescently labeled and their dynamics were directly visualized using fluorescence microscopy. Over a few hours, lateral association of filaments promoted formation of 880-nm-thick 2D colloidal membranes with an average lateral size of tens of microns. Shorter *litmus* rods were dissolved in such colloidal membranes (Fig. 1d, e). By viewing the membrane edge-on, we observed that isolated short rods aligned along the membrane normal, with one of the ends of each rod strongly anchored to the membrane-polymer interface and the other end located near the membrane midplane (Fig. 1f, g). The rods appeared effectively trapped on the interface. On rare occasions they would overcome the midplane energetic barrier to hop to the opposite side (Supplementary Movie 1). When viewed edge on the rods very quickly diffused away from the imaging plane, precluding us from quantifying their dynamics within a membrane.

These qualitative observations suggest that a short rod at the membrane midplane has a higher energy cost than when it is anchored at either interface. This can be rationalized by noting that each short rod end located in the membrane interior creates an effective dislocation defect, which increases the system distortion energy and excluded volume accessible to the depletant. The neighboring rods reduce the excluded volume by bending over an effective length scale to occupy the empty space created by a rod end, but this requires unfavorable bending energy. The importance of such defects and their effective interactions have been studied in the context of polymer nematic liquid crystals[35]. A short rod placed at the membrane midplane has two ends dissolved in the membrane and correspondingly generates two dislocation defects, whereas a short rod anchored at an interface creates only one defect and thus incurs a smaller free energy penalty (Fig. 1b, c).

Next, we assembled membranes using a virus mixture at higher ratio of short to long rods ($n_{mem}$=1) but the same depletant concentration. When viewed from above in a face-on configuration, such membranes appeared uniform in both the yellow and blue channels, indicating that both rods were uniformly dispersed throughout the membrane (Fig. 2c,e). However, when viewed edge-on, the membrane appeared different in the two channels. In the blue channel the membrane appeared uniform across its thickness, while in the yellow channel, two layers stacked on top of each other were clearly visible (Fig. 2b,d). These observations demonstrate that even at high concentrations, short rods preferentially dissolve in a membrane of long rods rather than in the depleting polymer. They also demonstrate that *fd-wt*/*litmus* membranes are simultaneously monolayers and bilayers. Under these condition the short rods are still preferentially anchored to the surface. It is possible, that a pair of short rods anchored at opposite interfaces effectively stack on top of each other (Fig. 2a). However, our imaging capabilities do not allow us to determine the fraction of short rods that have dimerized through end-to-end stacking as opposed to monomers that were previously visualized in very dilute regime.

Increasing the dextran concentration to 53 mg/mL increased the in-plane rods densities resulting in different phase behaviors. In this regime the membranes were no longer laterally uniform. Rather, we observed phase separated droplets enriched in the short rods, immersed in a background membrane enriched in long rods (Fig. 3a, b). To investigate the structures of such membranes we employed fluorescence, DIC, and LC-PolScope microscopy. First, when viewed in the edge-on configuration the entire membrane exhibited a bilayer structure (Fig. 3c). However, the droplets were much brighter, indicating that they were enriched in short rods. The system was in a dynamical equilibrium, as brightly labeled short rods were continuously exchanging between the enriched droplets and the background phase (Fig. 4a, Supplementary Movie 2). Second, when viewed with DIC microscopy, the bilayer droplets were barely visible. Contrast in DIC microscopy is generated by differences in the optical path length and the index of refraction, and is thus sensitive to different membrane thicknesses or in-plane densities. For instance, colloidal rafts that are 20% shorter than background membranes are easily visualized with DIC microscopy[20]. The poor visibility of bilayer droplets in DIC microscopy indicates a slight optical contrast between *litmus* droplets and the background membrane, confirming that droplets have a bilayer structure (Fig. 3d). Third, the droplets were not visible with the LC-PolScope technique, which is sensitive to local tilt away from the membrane normal (Fig. 3e)[36, 37]. In particular, LC-PolScope provides 2D spatial maps where the intensity of each pixel represents the magnitude and orientation of the local optical retardance. Hence, regions of the membrane where rods point perpendicular to the image plane appear dark due to their low birefringence, while regions where rods tilt away from the image plane are bright due to local birefringence. Therefore, the LC-PolScope measurements demonstrate absence of local twisting at the interface of bilayer droplets, in contrast to previously observed monodisperse colloidal rafts[20, 23].

We systematically varied the ratio of short to long rods ($n_{mem}$), while keeping the overall dextran concentration fixed at 53 mg/mL. Increasing the fraction of short rods $n_{mem}$ led to an increase in the average bilayer droplet size (Fig. 3f). We followed a large number of droplets at various area fractions over a period of days, and never observed even a single droplet coalescence event. This was the case even when the large droplets were almost touching each other (Fig. 3f), (Supplementary Movie 3,4). These observations suggest that the bilayer droplets are kinetically stabilized structures that have effective repulsive interactions. The bilayer rods initially formed from a few nuclei; these grew in size until the density difference between the droplet and the background reached the equilibrium value. Once this point was reached, droplets remained stable over four to five days. Their size did not significantly change over this time period, since coalescence did not occur. The absence of any coalescence events indicates the presence of repulsive interactions between droplets. Observing the samples on longer time scales revealed that the liquid-liquid coexistence of bilayer droplets with a monomer background becomes metastable with respect to a solid-liquid coexistence. Typically after 4 to 5 days we observed nucleation of a critical 2D crystal in the background long-rod membrane (Fig. 4b, Supplementary Movie 5). Subsequently, the crystalline phase grew over tens of minutes and the entire background monolayer membrane became solid. Following the dynamics of isolated short rods revealed that the bilayer droplets, coexisting with the solid background membranes, remained liquid-like (Supplementary Movie 6).

Increasing the ratio of short to long rods to $n_{nem}$=4, while keeping the dextran concentration fixed, led to formation of new structures. In particular, for these conditions we observed separate formation of thin *litmus* and thick *fd-wt* monolayer membranes. These different-thickness membranes frequently fused through lateral coalescence, enabling us to visualize the transition region where the membrane thickness changed from ~400 to ~880 nm (Fig. 5). Fluorescence microscopy, in which only short *litmus* rods were labeled, revealed a region where the membrane is 400 nm thick, and am adjacent dimmer region that is primarily composed of longer unlabeled *fd-wt* rods (Fig. 5a). These two regions were separated by a transition region marked by a thin much brighter layer of a defined width (Fig. 5b) (Supplementary Movie 7). This suggest that a short rod monolayer first transforms into a bilayer and this bilayer is fused to the longer rod membrane as shown in Fig. 5c. This hypothesis was supported by LC-PolScope images, which revealed that the *litmus* bilayer does not twist at the interface with the *fd-wt* monomer membrane as they have comparable length. In contrast, there was a strong structural anisotropy along the interface where the *litmus* bilayer transitions to a *litmus* monolayer (Fig. 5b). This transition was accompanied by twist of rods, leading to a strong in-plane birefringence signal. The transition region was easily visualized in DIC images since there was a large change in the optical path length due to the change in the membrane thickness.

To determine the region of phase space where each of the structures described above is stable, we systematically changed the two parameters that control the structure of long-short rod membranes, namely the depletant concentration and the ratio of short to long rods, $n_{mem}$ (Fig. 6). At low depletant concentrations viruses assembled into 3D tactoids, while at high depletant concentration they formed a 3D smectic phase comprised of stacks of membranes[38]. At intermediate concentrations the formation of 2D colloidal membrane was favored[17]. Within this regime, at lower *litmus* number fraction, short rods remained homogenously mixed with the background membrane. while increasing $n_{mem}$ lead to phase separation of short and long rods. For very large values of $n_{mem}$, we observed formation of distinct thin membranes composed of *litmus* virus and thick membranes composed of *fd-wt* virus.

Previous work demonstrated that the chirality plays an important role in stabilizing colloidal rafts of finite size[20-22]. To elucidate the role of chirality on formation of bilayer droplets, we examined a binary mixture composed of *litmus* and *fd*-Y21M. In contrast to *litmus* and *fd-wt*, *fd*-Y21M with a 6 nm diameter a contour length of 880 nm and a persistence length of 9.9 mm, forms right-handed cholesteric liquid crystal[39]. We found that changing the chirality of the longer rods does not appreciably influence the phase behavior. Fluorescent images indicate formation of *litmus* bilayers droplet floating in a *fd*-Y21M monolayer membranes (Fig. 7a). Although the two viruses have opposite chirality, LC-Polscope images indicate a lack of local twist along the interface. This suggests that dimerization and droplet formation of short rods is dominated by excluded volume interactions and is largely independent of the chirality. In colloidal membranes the rod twist necessarily couples to the local changes in the membrane thickness. If two rods have the same length but opposite chirality they cannot twist without creating local changes in the membrane thickness, which increases the effective surface tension of the membrane-depletant interface and thus costs energy.

The interface between a bilayer droplet and the background membrane exhibits pronounced fluctuations that provide additional evidence for the absence of any interfacial twist (Supplementary Movie 8)[40]. To analyze such fluctuations we assembled membranes consisting of phase separated fluorescently labeled *litmus* rods and unlabeled *fd-wt* rods. Analysis of a series of uncorrelated images of such an interface yielded the fluctuation spectrum $<a_q^2>$ as a function of wavevector $q$ (see Supplementary Information). Previously studied fluctuations of the exposed membrane edge were described with the following form: $<a_q^2> \sim \frac{k_b T}{\gamma + \kappa q^2}$ [41, 42], where $\gamma$ is the surface tension which dominates fluctuations at large wavelength (small $q$) values, and $\kappa$ is the bending elasticity that arises because rods near the edge twist away from the membrane normal, thus generating in-plane liquid crystalline order[19, 41]. The fluctuation spectrum of the interface between membrane bilayer and monolayer does not exhibit an asymptotic $1/q^2$ regime (Fig. 7c). This suggests that interfacial twisting is absent, which is consistent with the LC-Polscope images. The magnitude of the fluctuation spectrum at low $q$ yields provides an estimate of the interfacial tension ~57 $k_b T/\mu m$. Intriguingly, with increasing wavenumber $q$ the fluctuation spectrum does not remain flat but scales $\sim 1/q^4$, indicating the emergence of new physics at small separations. This is in contrast to molecular systems, where experimental measurements suggest that the surface tension decreases at small length scales, leading to enhanced fluctuations[43].

We have repeated similar analysis for the *litmus-fd* Y21M interface (Fig. 7b, Supplementary Movie 9). At a low volume fraction of short rods, *litmus* wets the membrane edge forming two closely adjoining interfaces, the outer membrane edge and the interface between the *litmus* bilayer and the *fd*-Y21 monolayer. The fluctuations of the outer membrane edge are consistent with previous studies, while the fluctuations of the inner interface has the same $q$ dependence as the previously studied *litmus/fd-wt* interface (Fig. 7c). This provides additional support for our previous observation that chirality couples to membrane thickness and therefore cannot influence the phase behavior of binary colloidal membranes in which there is no significant change in height.

**Theoretical model:** A simplified theoretical analysis elucidates the molecular forces that drive lateral phase separation of short rods within the membrane. We consider a mixture of two-rod species, with total area fractions $\phi_s$ for short rods and 1- $\phi_s$ for long rods. Above a critical area fraction, $\phi_s$*, the system phase separates into dense droplets enriched in short rods that coexist with a background membrane enriched in the long rods. For a depletant radius large compared to the rod radius, the osmotic free energy penalty for placing a long rod in the droplets phase is larger than that for a shorter rod mixing in the background phase, and thus we assume that the dense phase contains pure short rods[21]. We further assume that the area fraction of short rods in the background membranes, $\phi_s$, remains sufficiently low that we can neglect interactions between short rods.

The dependence of $\phi_s$* on the depletant concentration and the length difference between a short-rod dimer and a long rod ($\Delta L$) can then be obtained from a standard treatment of the thermodynamics of phase coexistence, at which the chemical potential of short rods in the two phases must be equal[21]. Since we assume that the dense phase consists of pure short rods, the chemical potential difference between the droplets and background membranes is given by $\delta\mu = k_B T \log(\phi_s) + \delta G$ where $\delta G$ is the free energy change associated with moving a short rod from the short rod bilayer into the background of long rods, to be calculated below. Thus, the critical area fraction is given by $\phi_s$*=exp($-\delta G/k_B T$). For a total area fraction of dimers below the coexistence area fraction, $\phi_s < \phi_s$*, the membrane remains homogeneous. Above the coexistence area fraction, the short-rod area fraction in the background phase is given by $\phi_s = \phi_s$*, with all remaining rods found in the raft phase.

To estimate $\delta G$, we consider the change in free energy associated with moving a short rod from the dimer bilayer into the long-rod background. If the long rods were perfectly rigid, there would be an increase in excluded volume $\Delta V \sim \Delta L \rho_{2D}$, with $\rho_{2D}$ the areal density and $\Delta L$=110 nm the height increase of the long rod monolayer over the short-rod bilayer. Thus, the free energy change would increase by $\delta G \sim \Pi \Delta V$ with $\Pi$ the osmotic pressure. However, the long rods have finite persistence length, and can deform to fill in some of this volume. Consider an isolated short rod aligned along the membrane composed of long rods, with one short-rod end anchored to the membrane-polymer interface, and the other end located near the membrane mid-plane (Fig. 1c). The long rods will then deform around the short-rod end near the mid-plane. To estimate the free energy associated with this deformation, we follow previous treatment of a chain end within a nematic of semiflexible filaments[44]. Filling the gap created by a chain end requires neighboring rods to deform over a distance $d$, the spacing between rods that is related to $\rho_{2D}$ by $d^2 = 2/(\rho_{2D}\sqrt{3})$. The distance along the contour of a long rod required for such a deformation to occur under the thermal energy $k_BT$ is given by the "deflection length", $l_d = (2d)^{2/3} l_p^{1/3}$ with $l_p = 2.8$ μm the persistence length of the long rods. Thus, there is an open space, or "shadow volume" in the vicinity of the chain end given by: $v_{\text{shadow}} \approx \gamma (\pi d^2/4) l_d$, with $\gamma$ a geometrical factor. Assuming the shadow volume has the shape of a cone, $\gamma = 1/3$. The free energy associated with the chain is then given by $\delta G = \Pi v_{\text{shadow}} - l_d F_{\text{int}}(d)$, where $F_{\text{int}}(d)$ is the interaction free energy per length due to electrostatics between pairs of rods separated at distance $d$. The latter term accounts for the fact that the open space of the shadow volume reduces interactions of the surrounding rods. Because the electrostatic screening length is $\kappa^{-1} \approx 1$ nm at the experimental conditions, the interaction between two rods separated by a distance $\sim 2d$ across the shadow volume can be taken as zero.

To estimate $F_{\text{int}}$ we neglect bending fluctuations of the rods and assume that the local concentration of counterions is equal to the bulk density, so that the local screening length is $\kappa^{-1} = 1$ nm. We relax both of these approximations in a forthcoming study in which we measure the equation of state of colloidal membranes[45]. However, relaxing these approximations does not significantly change our estimate of $\delta G$, so we retain them for simplicity. The electrostatic interactions between pairs of parallel rigid rods can be written as

$$\frac{F_{\text{pair}}(d)}{k_B T} = 2 \frac{\xi^2}{l_B} K_0(\kappa d) \approx \frac{\sqrt{2\pi} \xi^2 e^{-\kappa d}}{l_B \sqrt{\kappa d}}$$

where $K_0$ is the zero order modified Bessel function of the second kind, $l_B$ is the Bjerrum length, and $\xi = l_B \nu$ is the dimensionless effective charge density, with $\nu$ the effective linear charge density of the *fd* virus[46]. To calculate $\nu$ we note that counting the charges in the capsid protein and DNA yield a bare charge density of $\nu_0 \approx -7e / \text{nm}$ [34]. Next, we account for charge renormalization by counterions as described elsewhere[47, 48]. We note that the Debye-Huckel approximation accurately describes the form of the electrostatic potential in the far field, but over-predicts the potential in the near field. Therefore, we find the effective charge density for which the far-field potential is correct. We use an approximate analytical solution to the nonlinear Poisson Boltzmann equation around a cylinder [49], which matches a near-field solution to the Debye-Huckel far-field. Equating the far-field result to Eq. 3 then gives the effective charge density, as a function of the bare charge density $v_0$, screening length $\kappa$, and cylinder diameter $a$. For $v_0 = -7e / \text{nm}$, $\kappa^{-1} = 1$ nm, $l_B = 0.71$ nm in water, and diameter of $fd$ $a$=6.6nm, we obtain $\xi = 36.12$. Although the membrane has liquid in-plane order for polymer concentrations below 55 mg/ml, we make the simplifying assumption that the rods have local hexagonal order. Then the interaction free energy per unit length is given by $F_{\text{int}}(d_{\text{eq}}) = 3F_{\text{pair}}(d_{\text{eq}})$ with $d_{\text{eq}}$ the equilibrium lattice spacing, 3 interactions per rod (avoiding double-counting), and neglecting interactions beyond nearest neighbors due to screening.

Next, we need to estimate the equilibrium lattice spacing $d_{\text{eq}}$ as a function of the applied osmotic pressure $\Pi$. For consistency, we maintain the same level of approximation used to estimate $F_{\text{int}}$. At the equilibrium spacing the internal pressure from rod interactions will balance the applied osmotic pressure, $\Pi_{\text{int}}(d_{\text{eq}}) = \Pi$, with the internal pressure for hexagonally ordered rods given by $\Pi_{\text{int}} = -\frac{1}{\sqrt{3}d} \frac{\partial F_{\text{int}}}{\partial d}$[47, 50]. This results in an expression for the equilibrium spacing that can be solved numerically:

$$\Pi = k_\text{B} T \frac{\sqrt{6\pi\kappa}}{l_B d_{\text{eq}}^{3/2}} \xi^2 e^{-\kappa d_{\text{eq}}}.$$

Note that we assume that one rod end remains close to the membrane edge-plane because moving the rod into the middle of the membrane requires a second shadow volume, whose unfavorable free energy would outweigh the favorable increase in mixing entropy. There is also a driving force for a second short rod to "dimerize" with the first short rod, forming a chain of 2 rods aligned with the membrane normal, since this would require only one shadow volume. However, this effect is the same physics that drives phase separation, so this reaction will not become favorable until $\phi_s \geq \phi_s^*$. In other words, below $\phi_s^*$ the rods will only transiently dimerize, and will form permanent dimers at areal fractions comparable to those where they also bulk phase separate. We assume that the concentration of transient dimers is negligible.

If we explicitly account for the bending energy of neighboring rods and minimize the total free energy as a function of the length $l_d$, $\delta G$ decreases by a negligible amount. Similarly, we obtain a comparable yet independent estimate of $\delta G$ by calculating the free energy associated with a volume fluctuation of size $v_{shadow}$ according to the Gaussian model for particle density fluctuations[51-53], $\delta G = v_{shadow} / 2\kappa_T - l_d F_{int}(d)$ with $\kappa_T$ the isothermal compressibility estimated to be 6.3 mPa and 5.3 mPa at 50 and 55 mg/mL dextran, from the measurements of membrane density as a function of dextran concentration[45].

To calculate the critical ratio of short rods at which phase separation takes place, $n_{mem}^* = \phi_s^*/(1-\phi_s^*)$, as a function of the dextran concentration, we have used the raft density $\rho_{2D}$ as a function of dextran concentration measured using microfluidic technology[45], and a modification of the empirically measured virial expansion for dextran osmotic pressure: $\Pi = 0.0655\ (m_{W0}/m_W)c + 10.38c^2 + 75.3c^3$, with $c$ the dextran weight fraction, $\Pi$ the osmotic pressure in ATM, $m_W = 6.7 \times 10^5$ g/mol the dextran molecular weight in our experiments, and $m_{W0} = 3.7 \times 10^5$ g/mol the dextran molecular weight used for the measurement[54]. The osmotic pressure is relatively insensitive to molecular weight at these parameters[54-56]; we used the term $m_{W0}/m_W$ to correct the van't Hoff coefficient for the molecular weight difference. Dextran is non-ideal at the experimental concentrations, with $\Pi$ exceeding the van't Hoff result by a factor of 20.

We compared the theoretical prediction for the critical area fraction against experimental measurements at two dextran concentrations (Fig. 8). While two data points are not sufficient to test the accuracy of the theory, we observe close agreement at these two points. In further qualitative support of the theory, the measured critical area fraction where bulk phase separation takes place decreases with increasing dextran concentration, and the area fraction of short rods in the dilute phase $\phi_s$ is approximately independent of the total area fraction $\phi_{sT}$ within the coexistence region of the phase diagram.

Despite the agreement between theory and experiment, we note that a similar calculation of rod interactions based on electrostatics overestimated experimentally measured values by a factor of 3-5, suggesting there could be a cancellation of errors. Thus, we note approximations that we have made. Firstly, we have neglected rod bending fluctuations; these quantitatively increase $\Pi_{int}$ and $F_{int}$ [45], but will not qualitatively change the results (to some extent the error in $\Pi_{int}$ and $F_{int}$ cancels). Secondly, we have neglected the unfavorable entropic penalty due to suppression of rod protrusions required for the stacking of two smectic layers that occurs in the raft domain[17, 39, 57], and we have not accounted for the finite concentration of long rods found in the dimer phase. We also neglect the entropy associated with the fact that the two layers within the *litmus* bilayer are free to slide past each other. This is reminiscent of the entropic considerations that drive the transition from the smectic to columnar phase observed in hard rods at high concentrations[58]. However, this entropic contribution would diminish with increasing depletant concentration, in contrast to the experimentally observed trend for $\phi_s^*$, and thus is not dominant. Finally, we found that accounting for the different surface charges of *fd* and *litmus* has a negligible effect on our estimate of the critical area fraction.

**Discussion and Conclusions:** Previous experiments have demonstrated that rods with opposite chirality and a length difference of 30% robustly assemble into highly uniform micron-sized colloidal rafts[20-23]. In comparison, here we study the phase behavior of rod-like inclusions that are approximately half the length of the host membrane. We demonstrate that such rods dissolved in an anisotropic environment of a colloidal membrane robustly anchor to the membrane-polymer interface. Our observations of anchoring are qualitative, and quantifying surface anchoring in colloidal membrane is challenging, since in any field of view one observes few if any edge-on short rods, and rods that are observed quickly diffuse out of the image plane. Furthermore, the colloidal membranes themselves fluctuate, and quantifying the dynamics of short rod requires dual labeling of the entire membrane as well as the isolated rods.

The surface anchoring effect described here might be relevant in other contexts. For example, similar considerations could play a role in the structure and dynamics of smectic liquid crystals comprised of semi-flexible filaments[59]. Shorter rods dissolved in such a system should anchor to the smectic layer edge in order to reduce the entropic penalty due to dislocation defects . This prediction can be experimentally tested, as the quantification of single rod dynamics in *fd* nematic liquid crystals has been extended to smectic liquid crystals[25, 60, 61]. Recent work also showed that rods that are slightly longer than the smectic layer exhibit faster diffusion. The dynamics of short surface anchored rods should be more easily quantified in bulk smectics, in comparison to colloidal membranes, since for bulk smectics one can observe a full field of edge-on layers, thus enabling better statistics. Furthermore, in comparison to colloidal membranes smectic phases do not fluctuate on optical length scales, and thus one only needs to track the short-rod dynamics. It is likely that the filament stiffness controls the strength of the anchoring, as the more rigid rods with Y21M mutation would heal from dislocations over longer distances and thus incur a larger entropic penalty.

At higher concentrations short rods more effectively occupy space by dimerizing and phase separating from the host membranes, which lowers the entropy of the depleting polymers that envelop the colloidal membranes. The uniformly mixed binary bilayer colloidal membranes are similar to the smectic-A2 phase that has been observed in molecular liquid crystals[62, 63]. Such phases have also been theoretically predicted for a suspension of hard rods[64-66], but have not yet been seen in experiments. It should be feasible to search for such phenomena either using filamentous viruses or colloidal silica spheres, as both these systems robustly form smectic phases[59, 61, 67]. We note that our current imaging techniques do not allow us to determine the point at which isolated surface-anchored rods dimerize, and the simple theoretical arguments described previously suggest that dimerization takes place at the same volume fraction as the bulk phase separation.

The lack of the bilayer droplets coalescence is more challenging to explain. Previous work on binary colloidal membranes demonstrated that the twist surrounding each colloidal raft induces long-ranged repulsive interactions that suppress raft coalescence[20]. Twist couples to the local changes in the membrane thickness. Without incurring additional energetic cost in surface tension, twist at the droplet edge can only develop for rods of different lengths. Dimerizing *litmus* rods have an effective length that is comparable to that of the background membrane. Consequently, there is no measurable interfacial twist as is evidenced by quantitative LC-PolScope microscopy. In the absence of edge bound twist, the exact mechanism that suppresses lateral coalescence of bilayer droplets remains unknown.

**Acknowledgment:** We acknowledge useful discussions with Greg Grason. We also acknowledge support of National Science Foundation through grants: NSF-MRSEC-1420382 and NSF-DMR-1609742. We also acknowledge use of Brandeis MRSEC optical and biomaterial synthesis facilities supported by NSF-MRSEC-1420382.

# Reference List

1. J. N. Israelachvili, *Intermolecular and surface forces*, Academic press, 2011.
2. P. Poulin, H. Stark, T. Lubensky and D. Weitz, *Science*, 1997, **275**, 1770-1773.
3. I. Muševič, M. Škarabot, U. Tkalec, M. Ravnik and S. Žumer, *Science*, 2006, **313**, 954-958.
4. H. Mundoor, B. Senyuk and I. I. Smalyukh, *Science*, 2016, **352**, 69-73.
5. Y. Lin, H. Skaff, T. Emrick, A. Dinsmore and T. P. Russell, *Science*, 2003, **299**, 226-229.
6. M. Cavallaro, L. Botto, E. P. Lewandowski, M. Wang and K. J. Stebe, *Proceedings of the National Academy of Sciences*, 2011, **108**, 20923-20928.
7. J.-C. Loudet, A. M. Alsayed, J. Zhang and A. G. Yodh, *Physical review letters*, 2005, **94**, 018301.
8. B. Madivala, J. Fransaer and J. Vermant, *Langmuir*, 2009, **25**, 2718-2728.
9. P. Pieranski, *Physical Review Letters*, 1980, **45**, 569.
10. S. Marčelja, *Biochimica et Biophysica Acta (BBA)-Biomembranes*, 1974, **367**, 165-176.
11. S. L. Veatch and S. L. Keller, *Biochimica et Biophysica Acta (BBA)-Molecular Cell Research*, 2005, **1746**, 172-185.
12. T. Baumgart, S. T. Hess and W. W. Webb, *Nature*, 2003, **425**, 821-824.
13. B. J. Reynwar, G. Illya, V. A. Harmandaris, M. M. Müller, K. Kremer and M. Deserno, *Nature*, 2007, **447**, 461-464.
14. D. Lingwood and K. Simons, *Science*, 2010, **327**, 46-50.
15. J. Israelachvili, S. Marčelja and R. G. Horn, *Quarterly reviews of biophysics*, 1980, **13**, 121-200.
16. E. Barry and Z. Dogic, *Proc Natl Acad Sci U S A*, 2010, **107**, 10348-10353.
17. Y. S. Yang, E. Barry, Z. Dogic and M. F. Hagan, *Soft Matter*, 2012, **8**, 707-714.
18. M. Siavashpouri, C. H. Wachauf, M. J. Zakhary, F. Paetorius, H. Dietz and Z. Dogic, *Nature Materials*, 2017.
19. L. Kang, T. Gibaud, Z. Dogic and T. C. Lubensky, *Soft Matter*, 2016, **12**, 386-401.
20. P. Sharma, A. Ward, T. Gibaud, M. F. Hagan and Z. Dogic, *Nature*, 2014, **513**, 77-+.
21. L. Kang and T. C. Lubensky, *Proceedings of the National Academy of Sciences*, 2017, **114**, E19-E27.
22. R. Sakhardande, S. Stanojeviea, A. Baskaran, A. Baskaran, M. F. Hagan and B. Chakraborty, *Physical Review E*, 2017, **96**, 012704.
23. J. Miller, C. Joshi, P. Sharma, A. Baskaran, G. M. Grason, M. F. Hagan and Z. Dogic, *arXiv preprint arXiv:1902.03341*, 2019.
24. J. Sambrook, E. F. Fritsch and T. Maniatis, *Molecular cloning: a laboratory manual*, Cold spring harbor laboratory press, 1989.
25. M. P. Lettinga, E. Barry and Z. Dogic, *Europhysics Letters*, 2005, **71**, 692-698.
26. C. P. Brangwynne, G. H. Koenderink, E. Barry, Z. Dogic, F. C. MacKintosh and D. A. Weitz, *Biophysical Journal*, 2007, **93**, 346-359.
27. A. Lau, A. Prasad and Z. Dogic, *EPL (Europhysics Letters)*, 2009, **87**, 48006.
28. R. Oldenbourg, *Journal of Microscopy-Oxford*, 2008, **231**, 419-432.
29. J. Canny, *IEEE Transactions on Pattern Analysis and Machine Learning* 1986, **8(6)**, 679–698
30. Z. Dogic and S. Fraden, *Philosophical Transactions of the Royal Society of London A: Mathematical, Physical and Engineering Sciences*, 2001, **359**, 997-1015.
31. K. R. Purdy and S. Fraden, *Phys Rev E Stat Nonlin Soft Matter Phys*, 2004, **70**, 061703.
32. E. Grelet and S. Fraden, *Physical review letters*, 2003, **90**, 198302.
33. F. Tombolato, A. Ferrarini and E. Grelet, *Physical review letters*, 2006, **96**, 258302.
34. G. Abramov, R. Shaharabani, O. Morag, R. Avinery, A. Haimovich, I. Oz, R. Beck and A. Goldbourt, *Biomacromolecules*, 2017.
35. J. V. Selinger and R. F. Bruinsma, *Journal de Physique II*, 1992, **2**, 1215-1236.
36. R. Oldenbourg and G. Mei, *Journal of microscopy*, 1995, **180**, 140-147.
37. E. Barry, Z. Dogic, R. B. Meyer, R. A. Pelcovits and R. Oldenbourg, *The Journal of Physical Chemistry B*, 2008, **113**, 3910-3913.

38. Z. Dogic, *Physical review letters*, 2003, **91**, 165701.
39. E. Barry, D. Beller and Z. Dogic, *Soft Matter*, 2009, **5**, 2563-2570.
40. D. G. Aarts, M. Schmidt and H. N. Lekkerkerker, *Science*, 2004, **304**, 847-850.
41. T. Gibaud, E. Barry, M. J. Zakhary, M. Henglin, A. Ward, Y. Yang, C. Berciu, R. Oldenbourg, M. F. Hagan, D. Nicastro, R. B. Meyer and Z. Dogic, *Nature*, 2012, **481**, 348-351.
42. L. L. Jia, M. J. Zakhary, Z. Dogic, R. A. Pelcovits and T. R. Powers, *Physical Review E*, 2017, **95**, 060701.
43. C. Fradin, A. Braslau, D. Luzet, D. Smilgies, M. Alba, N. Boudet, K. Mecke and J. Daillant, *Nature*, 2000, **403**, 871.
44. A. V. Tkachenko, *Physical review letters*, 1996, **77**, 4218.
45. A. J. Balchunas, R. A. Cabanas, M. J. Zakhary, T. Gibaud, S. Fraden, P. Sharma, M. Hagan and Z. Dogic, *arXiv preprint arXiv:1905.13227*, 2019.
46. S. L. Brenner and V. A. Parsegian, *Biophys. J.*, 1974, **14**, 327-334.
47. T. Odijk, *Biophys. Chem.*, 1993, **46**, 69-75.
48. A. Stroobants, H. N. W. Lekkerkerker and T. Odijk, *Macromolecules*, 1986, **19**, 2232-2238.
49. J. R. Philip and R. A. Wooding, *The Journal of Chemical Physics*, 1970, **52**, 953-959.
50. R. Podgornik and V. A. Parsegian, *Macromolecules*, 1990, **23**, 2265-2269.
51. G. Graziano, *Chemical Physics Letters*, 2007, **446**, 313-316.
52. D. M. Huang and D. Chandler, *Physical Review E*, 2000, **61**, 1501.
53. G. Hummer, S. Garde, A. Garcia, M. E. Paulaitis and L. R. Pratt, *J. Phys. Chem.*, 1998, **102**, 10469-10482.
54. H. Vink, *Eur. Polym. J.*, 1971, **7**, 1411-&.
55. V. Parsegian, R. Rand, N. Fuller and D. Rau, in *Methods in enzymology*, Elsevier, 1986, vol. 127, pp. 400-416.
56. F. Vérétout, M. Delaye and A. Tardieu, *Journal of molecular biology*, 1989, **205**, 713-728.
57. Y. Yang and M. F. Hagan, *Physical Review E*, 2011, **84**, 051402.
58. P. Bolhuis and D. Frenkel, *The Journal of chemical physics*, 1997, **106**, 666-687.
59. Z. Dogic and S. Fraden, *Physical review letters*, 1997, **78**, 2417.
60. L. Alvarez, M. P. Lettinga and E. Grelet, *Physical review letters*, 2017, **118**, 178002.
61. M. P. Lettinga and E. Grelet, *Physical review letters*, 2007, **99**, 197802.
62. G. Sigaud, F. Hardouin, M. Achard and H. Gasparoux, *Le Journal de Physique Colloques*, 1979, **40**, C3-356-C353-359.
63. K. Okoshi, A. Suzuki, M. Tokita, M. Fujiki and J. Watanabe, *Macromolecules*, 2009, **42**, 3443-3447.
64. T. Koda and H. Kimura, *Journal of the Physicsl Society of Japan*, 1994, **63**, 984-994.
65. R. P. Sear and G. Jackson, *The Journal of chemical physics*, 1995, **102**, 2622-2627.
66. G. Cinacchi, L. Mederos and E. Velasco, *The Journal of chemical physics*, 2004, **121**, 3854-3863.
67. A. Kuijk, A. van Blaaderen and A. Imhof, *Journal of the American Chemical Society*, 2011, **133**, 2346-2349.

**Figures:**

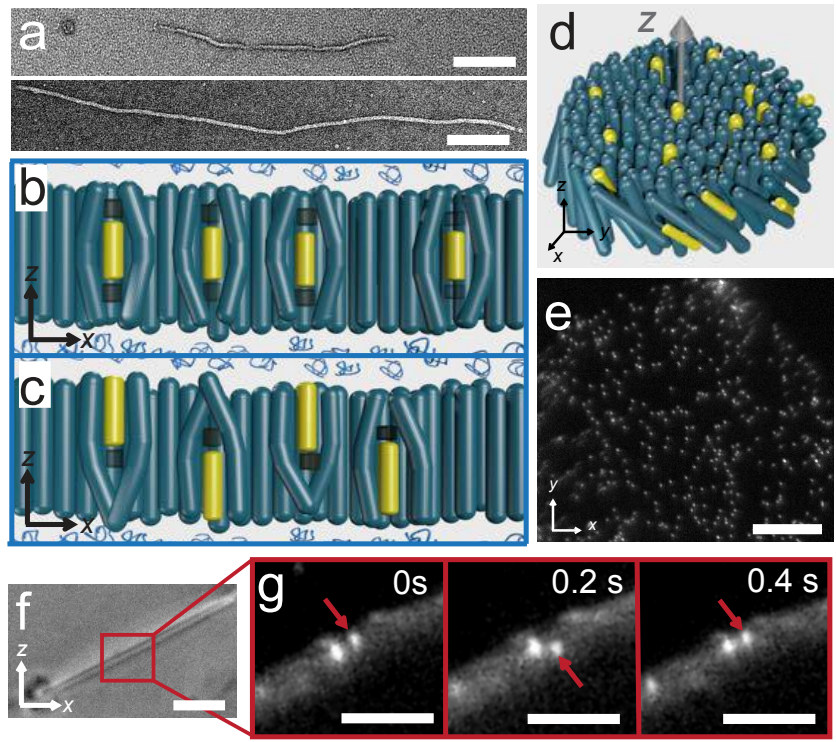

**Figure 1. Short rods dissolved in a colloidal membrane anchor to the membrane surface. a)** Transmission electron micrographs of 385-nm-long *litmus* and 880-nm-long *fd-wt* bacteriophages. Scale bars, 100 nm. **b)** Edge-on schematic of a colloidal membrane consisting of *fd-wt* monomers, in which *litmus* rods preferentially dissolve in the membrane midplane. This configuration generates entropically unfavorable void volumes above and below short rods end(*Litmus*: yellow, *fd-wt*: blue), and is not observed in experiments. Dark regions represent the excess of empty space that is inaccessible to depletant polymers. **c)** Edge-on schematic of an entropically favorable configuration of the membrane, in which short rods are anchored to the membrane-polymer interface to reduce the free volume of the system. **d)** Schematic of a self-assembled binary membrane consisting of long *fd-wt* (blue) and dilute short *litmus* rods (yellow). **e)** Face-on fluorescence image of a homogenously mixed membrane composed of *litmus* dimers and *fd-wt* monomers, demonstrating that short *litmus* rods are uniformly dispersed throughout the membrane. *litmus* is fluorescently labeled. Scale bar, 5 μm. **f)** Edge-on DIC image of a similar membrane. Scale bar, 2 μm **g)** Fluorescence image of a membrane viewed edge-on shows that short rods are anchored to the membrane-polymer interface. Infrequently they are observed to hop between the opposite interfaces within a fraction of a second. Scale bars, 1 μm.

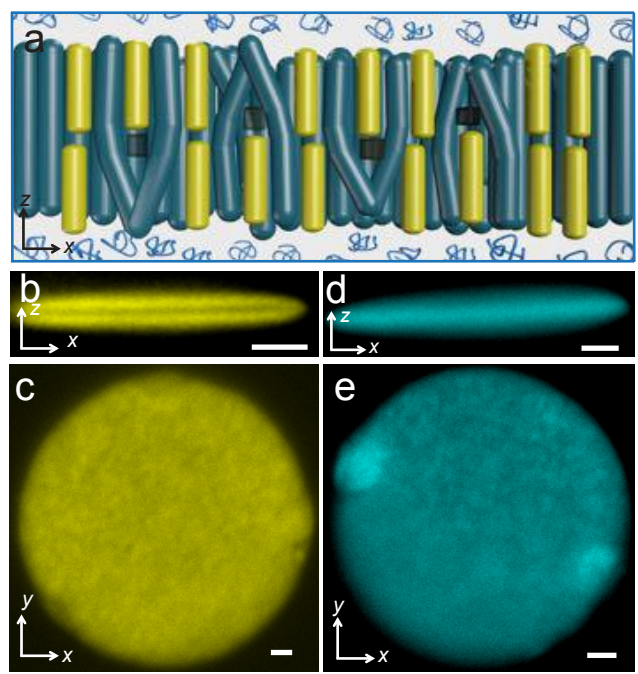

**Figure 2. Binary *fd-wt/litmus* membranes have a laterally uniform composite bilayer-monolayer structure at low depletant concentration. a)** Schematic of a uniformly mixed binary colloidal membrane at dextran concentrations of 40 mg/mL. **b)** Fluorescence image of a binary colloidal membrane viewed in the yellow channel, revealing that *litmus* virus is organized into a bilayer structure. **c)** Similar membrane in a face-on configuration, revealing a laterally uniform membrane. **d)** Fluorescence image of a binary colloidal membrane viewed in the blue channel, revealing that *fd-wt* virus form a monolayer membrane. **e)** Similar membrane in face-on configuration. Scale bars, **2** μm.

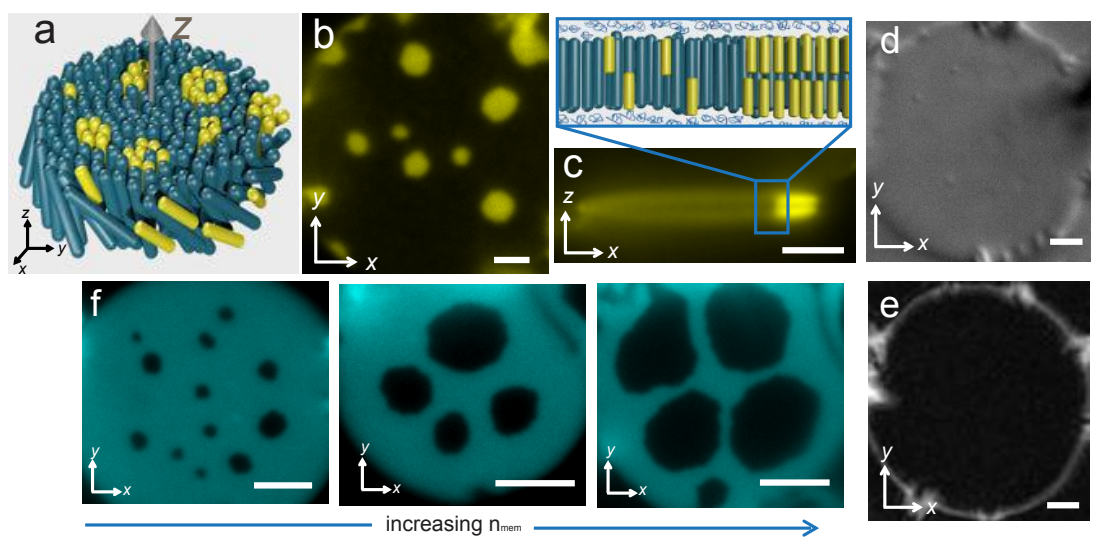

**Figure 3. Phase separated bilayer droplets coexist with the monolayer background membrane at high depletant concentration. a)** Schematic illustration of bilayer droplets enriched in short viruses that coexist with the background membrane enriched in long rods (*Litmus*: yellow, *fd-wt*: blue). **b)** Fluorescence image of a face-on membrane assembled at high depletant concentration (~53 mg/mL). Short rods are fluorescently labeled. **c)** Fluorescence image and schematics of a similar phase separated membrane viewed in an edge-on configuration. **d)** DIC micrograph of the phase-separated membrane, showing slight contrast along the droplet edge, which demonstrates that short rods form bilayers in the background of long monomer rods. **e)** LC-PolScope image of the membrane, demonstrating the absence of interfacial twist along the edge of bilayer droplets. **f)** Phase-separated membranes formed at increasing volume fractions of short rods increases the size of bilayer droplets. The ratio of *litmus* rods to *fd-wt* rods ($n_{mem}$) increases from 1 to 1.5, and to 2. *fd-wt* is fluorescently labeled (blue). The depletant concentration is 53 mg/mL. All scale bars, 5 μm.

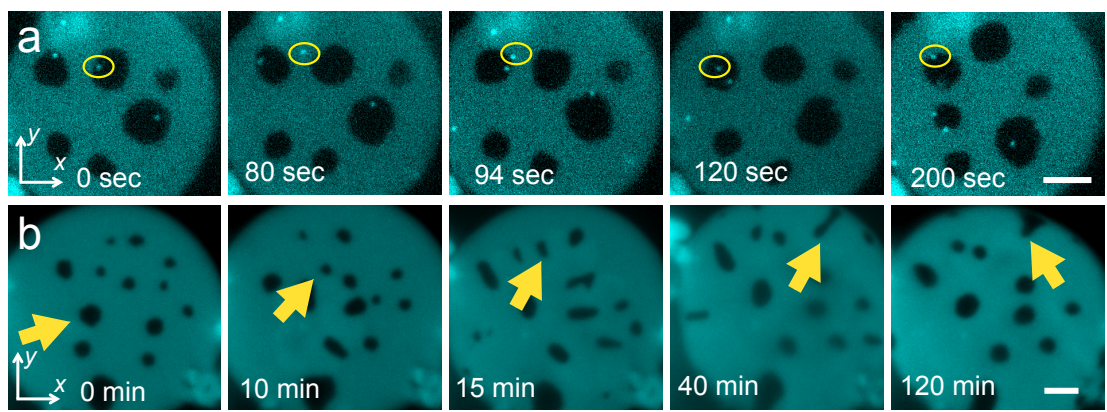

**Figure 4. Single-rod dynamics in a phase separated membrane. a)** Time-lapse image of a phase-separated 2D membrane consisting of fluorescently labeled *fd-wt* (blue) and unlabeled *litmus* rods at $n_{mem}$=1.5. A low fraction of highly labeled *Litmus* rods (bright blue, 1:10,000) are observed as bright points in the membrane. Tracking single rods demonstrates that they continuously exchange between the two coexisting phases. **b)** Time lapse of a 2D membrane in which the background phase crystallizes over a period of few hours. The membrane consists of fluorescently labeled *fd-wt* monomers (blue) and unlabeled *litmus* rods at $n_{mem}$=1 and depletant concentration is 55 mg/mL. The images have been taken 5 days after preparing the sample. The yellow arrow indicated the position of the solid-liquid interface during the growth pathway. All scale bars, 5 μm.

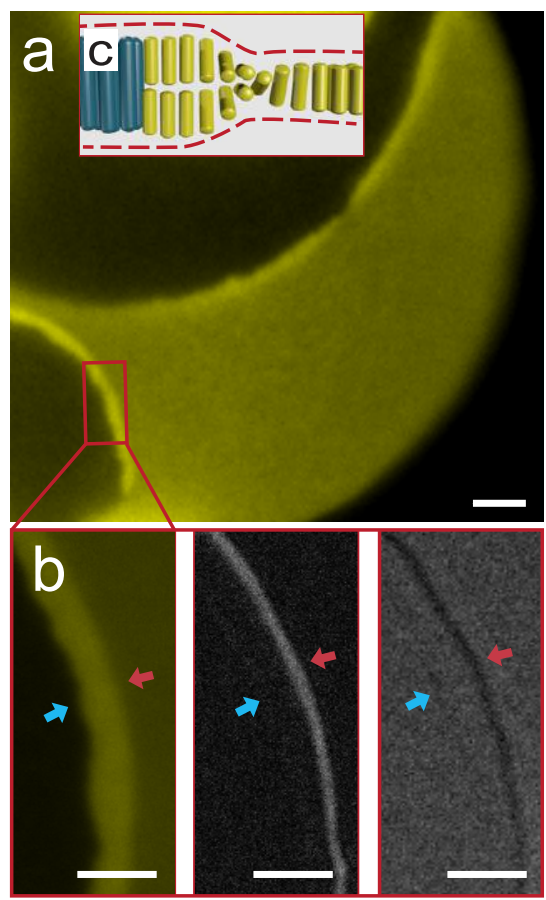

**Figure 5. Heterogeneous thickness membranes form at high fractions of short rods. a)** Membrane consisting of labeled *litmus* rods (yellow) exhibit demixed domains with spatially varying heights. The sample was prepared at $n_{mem}$ = 4 and depletant concentration 55 mg/mL. The dark region is ~880 nm thick membrane enriched in unlabeled *fd-wt*, and the light yellow region corresponds to a ~400 nm thick *litmus* monolayer. The bright yellow region indicates the transition regime with changing membrane thickness. Scale bar, 5 μm. **b)** Fluorescence, LC-PolScope, and DIC images of the transition region. Blue arrows indicate the interface between *fd-wt* monomers and *litmus* dimers ;LC-PolScope reveals a lack of twist (dark region) and DIC shows a small contrast along this interface. Red arrows indicates the transition from *litmus* bilayer to monomers. Measurable LC-Polscope and DIC contrast in this region indicates significant twist and changing membrane thickness. Scale bars, 2μm. **c)** Schematic of an edge-on view of a membrane shows the structure of the transition regime (*fd-wt*: blue, *litmus*: yellow).

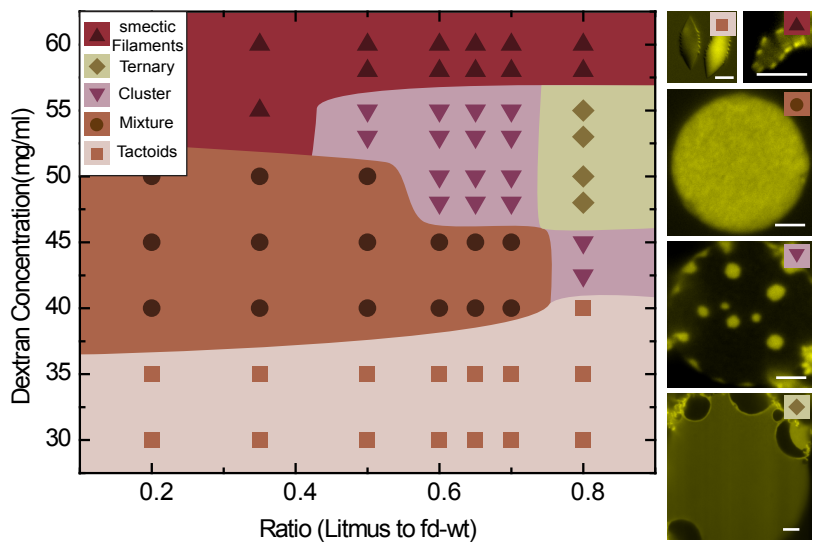

**Figure 6. Phase diagram of binary *fd-wt/litmus* colloidal membranes.** The phase diagram of is plotted as a ratio of long to short rods and the depletant concentration. The images show micrographs of different structures found in the phase diagram and their corresponding symbols: tactoids (squares), smectic filaments (upright triangles)**,** homogenously mixed membranes (circles), phase separated membranes (inverted triangles) and heterogeneous membranes (diamonds). *Litmus* rods are fluorescently labeled (yellow). The NaCl concentration is 135 mM; Scale bars, 5 μm.

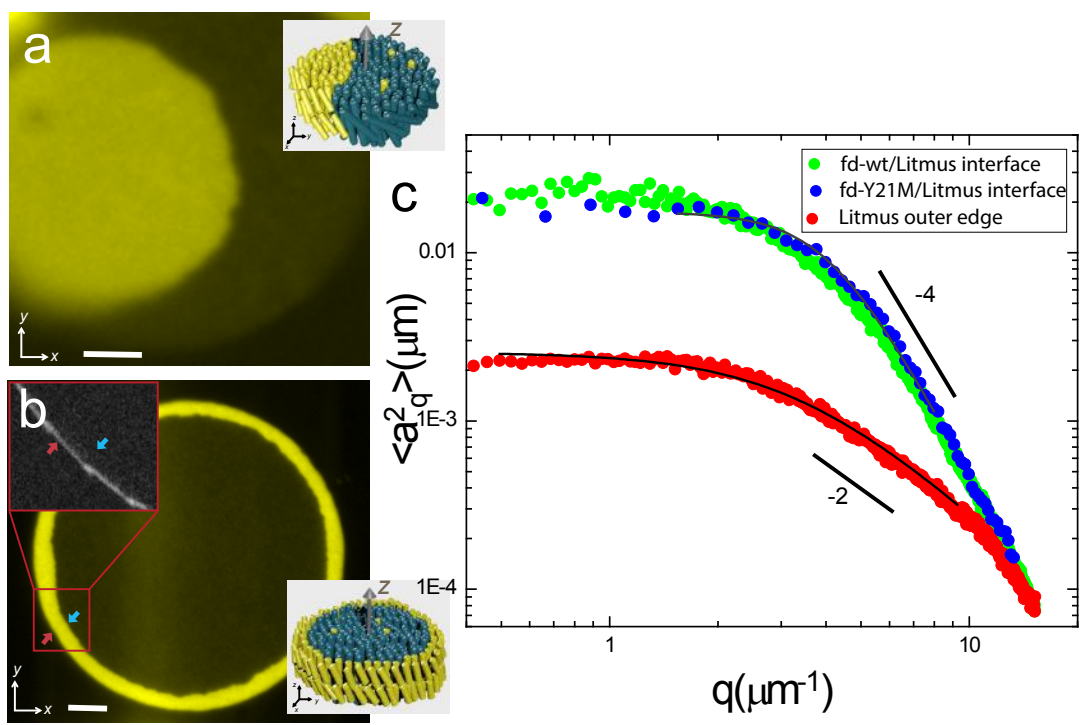

**Figure 7. Interfacial tension of the bilayer droplet. a)** Fluorescence image of a bulk phase-separated membrane at $n_{mem}$= 2.5 and depletant concentration 50 mg/mL. The schematic illustrates a bulk phase-separated membrane. **b)** Fluorescence image of a bulk phase-separated membrane comprised of right-handed *fd*-Y21M and left-handed *litmus*. $n_{mem}$= 2.5 and the dextran concentration is 53 mg/mL. *Litmus* bilayers wet the membrane edge. Inset: LC-Polscope image indicates that *litmus* dimers do not tilt at the inner *fd*-Y21M interface, but twist along the outer membrane edge. The schematic illustrates a bulk phase-separated membrane. **c)** Fluctuation spectrum of bilayer droplets dissolved in a *fd-wt* monolayer (green line), the inner edge of the *litmus*/*fd*Y21M interface (blue line), and the outer edge of the *fd*-Y21M/litmus membrane (red line). Dextran concentration is 53 mg/ml.. Scale bars, 5 μm.

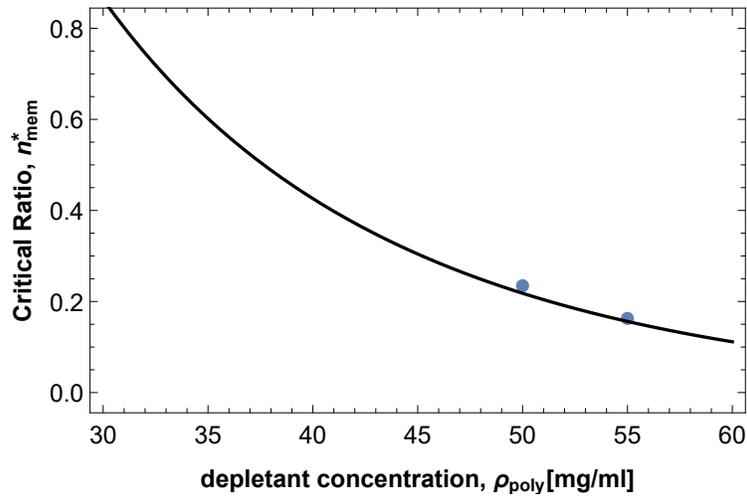

**Figure 8. Theoretical estimate of the critical ratio of short rods, $\phi_s^*$, above which phase separation will occur, as a function of dextran concentration.** The two experimental points are estimates of the ratio of short rods in the dilute phase at 50 mg/mL and 55 mg/mL. The experimentally measured ratio in the dilute phase is independent of dextran concentration above the phase-separation threshold.